\documentclass[a4paper,12pt]{article}

\usepackage{amssymb}
\usepackage{amsmath}
\usepackage[dvips]{graphicx}
\begin{document}

\vspace{1cm}

\begin{center} \LARGE \section* {Invariant variational principle\\ for
Hamiltonian mechanics}\end{center}

\vspace{1cm}

\begin{center} Alexey V. Golovnev$^1$, Alexander S. Ushakov$^2$\\

\end{center}
\vspace{1cm}
\indent
1. Ludwig-Maximilians University, astroparticle physics and cosmology department. Theresienstr. 37,
D-80333, Munich, Germany (on leave from
St.Petersburg State University).
Alexey.Golovnev@physik.uni-muenchen.de

2. St.Petersburg State University, V.A.Fock Institute of Physics. Ulianovskaya ul. 1, Petrodvoretz, 
Saint-Petersburg, Russia. asushakov@gmail.com
\vspace{1cm}
\begin{center} Abstract
\end{center}
{\it It is shown that the action for Hamiltonian equations of
motion can be brought into invariant symplectic form. In other words, it can be formulated directly in terms
of the symplectic structure $\omega$ without any need to choose some
1-form $\gamma$, such that $\omega= d \gamma$, which is not unique and 
does not even generally exist in a global sense.}
\vspace{1cm}

 {\LARGE \section {Introduction}}

Hamiltonian mechanics is defined by a smooth function
(Hamiltonian) on an even-dimensional manifold $M^{2n}$ equipped
with a symplectic structure, i.e. a closed non-degenerate differential
form,
 \begin{equation*}
\omega = \sum_{\mu > \nu} \omega_{\mu \nu}(x)dx^\mu\land dx^\nu=
\frac{1}{2}\sum_{\mu, \nu} \omega_{\mu \nu}dx^\mu\land dx^\nu,\quad  x\in M^{2n}.
\end{equation*}
 Using Darboux transformation these coordinates may be devided into two groups,
$x = (q^1, . . ., q^n, p_1, . . ., p_n)$ with $q^i$ and $p_i$
being generalized coordinates and momenta respectively, but this transformation is
not unique and in general can be performed only locally.

Equations of motion are given by Poisson brackets
\begin{equation}
\label{Poisson}
 \dot x^{\mu} = \left\{x,H\right\}=\sum^{2n}_{\nu = 1} \omega^{\mu\nu}
(x) \partial_\nu H,
\end{equation} where the right-hand side can be regarded as a vector obtained by
contraction of Poisson bivector (dual to $\omega$) with the
differential form $dH$; in local coordinates
 the bivector is represented by
$\omega^{\mu\nu}$, the matrix inverse to
 $\omega_{\mu\nu}$.  In the case of the standard symplectic
 structure, i.e.
 \begin{equation*}
\omega^{\mu\nu} = \left( \begin{array}{cc}
 O_n & E_n \\
 -E_n & O_n
 \end{array} \right),
 \end{equation*}
where $E_n$ and $O_n$ are the unit and the zero matrices respectively, a simple
action exists for these equations of motion \cite{Arnold}:
\begin{equation}
\label{action}
S= \int^{t_2}_{t_1} \left(\sum_{\alpha = 1}^{n}
p_\alpha \dot{q}^\alpha - H\right)\mbox{d}t,
\end{equation}
with the boundary conditions $q(t_1) = q_1 ,\, q(t_2) = q_2\,$.

But the expression (\ref{action}) does not seem satisfactory from geometric
point of view because it is suitable only in Darboux coordinates
 and does not use
the symplectic form explicitly. So, our goal would be to find an invariant
geometric form of the action. In Sections 2, 3 we discuss 
dynamical systems with exact symplectic forms and, first of all, the very important case of ``classical''
Hamiltonian systems (i.e. those in $\mathbb R^{2n}$ with
the standard symplectic structure). The special emphasis is placed on
the problem of boundary conditions \cite{alisa}. After that we explore non-exact symplectic forms (see also
\cite{McEwan} for K{\" a}hler manifolds as phase spaces, 
which are unitary phase spaces in the terminology of \cite{McEwan}).
To the best of our knowledge there is no invariant variational
principle in the literature for the general case. 
However, such systems may be of interest for different reasons. For example,
systems with gyroscopic forces can not be presented in a
straight-forward way as Hamiltonian systems with one-valued
Hamiltonians \cite{Surio}. The problem can be solved
\cite{Surio,Novikov} by some non-exact symplectic structure. 
In particular, the so-called Kirchhoff type systems \cite{Novikov} can be brought into Hamiltonian form on a cotangent bundle over 2-sphere with a volume form of the configuration space ${\mathcal S}^2$ being added to the standard symplectic structure. It is enough to mention that systems of this kind include \cite{Novikov} a rigid body in an ideal incompressible fluid at rest at infinity, a rigid body with a fixed point in axially symmetric potential field, spin dynamics in the A-phase of the superfluid ${}^3$He (Leggett equations), etc; and at the quantum level we would have a sphere with non-commutative coordinates. 
Motivated by these reasons, in Sections 4, 5 we establish an invariant variational principle
for arbitrary Hamiltonian systems. The last Section 6 is devoted to
some examples and discussions.\\

 {\LARGE \section {Brief review of known results}}

 As it was mentioned above, for classical Hamiltonian systems one can use the 
non-invariant variational principle with the
 action (\ref{action}) to obtain the equations of motion (\ref{Poisson}).
This action can be brought into invariant form as follows \cite{CoZeh,ser}:
 \begin{equation}
\label{actprime}
S= \int^{t_2}_{t_1} \left(
\frac{1}{2}\,
 \omega_{\mu\nu}\, \dot{x}^{\nu} x^{\mu} - H \right) dt =
 \int^{t_2}_{t_1}\left( \frac{1}{2}\,(p_i \dot{q}^i - \dot{p}_i
 q^i)- H \right) dt,
 \end{equation}
$ \mu, \nu = 1, \ldots, 2n,\quad i = 1, \ldots, n,$ where the
summation over repeated indices is assumed and Dirichlet boundary conditions
($\delta q(t_1)\!=\!\delta q(t_2)\!=\!\delta p(t_1)\!=\!\delta
p(t_2)\!=\!\!0$) on both coordinates and momenta variations are imposed, which
is two times as much as the number of possible independent
boundary conditions on the values of coordinates and momenta
themselves in Cauchy or boundary problem for the equations of motion.
We have a mismatch between the number of required variational boundary conditions
and the number of independent solutions of equations of motion. 

So, the Hamiltonian variational principle differs in this sense seriously from
the Lagrangian one. Actually, one may suspect that it can be formulated in a more usual way by
fixing the values of the functions $\phi_i=\arctan\frac{q_i}{p_i}$ at the boundary points.
Indeed, variations of $\phi$-s equal $$ \delta\phi_i =
\frac{p_i\delta q_i - q_i\delta p_i}{q_i^2+p_i^2};$$ and $\delta\phi_i=0$ at $t=t_0$ and
$t=t_1$ is exactly what we need to make the boundary terms in the first variation of 
(\ref{actprime}) equal zero. Equations $\phi_i=const$ define some
$n$-dimensional submanifolds in $2n$-dimensional phase space. And it would
be curious to introduce new pairs of coordinates in
the following way: $$q_i =r_i\sin\phi_i\quad p_i = r_i\cos\phi_i.$$
With the definition $P_i=r_i^{2}/2$ it
is easy to get $\frac{1}{2}(p_i\dot{q_i} - q_i\dot{p_i}) = P_i \dot{\phi_i}$ and
\begin{equation*}
 S= \int^{t_2}_{t_1} \left( \sum_i P_i\, \dot{\phi_i} -
H(\phi,P)\right) dt.
\end{equation*}
It looks quite good. But the point is that
these variables are nothing more than
a particular choice of Darboux coordinates. We could well have started with the action
(\ref{actprime}) for these canonical variables and considered other
prefered functions of the form $\tilde\phi_i=\arctan\frac{\phi_i}{P_i}$.
Hence we should stress once more that there is {\it no
invariant way} to fix only a half of boundary conditions without sticking to some
coordinate choice, because what we have done here
is just a canonical transformation to new variables, $P_i$ and $\phi_i$.

It is of crucial importance for the geometric meaning of the action (\ref{actprime}) that the phase space is
${\mathbb R}^{2n}$ because the radius-vectors are used. (But the symplectic form could
be arbitrary.) However, there is a natural analogue of the action (\ref{action}) which
is valid for any exact symplectic manifold. (Note still that for any compact symplectic
manifold (without boundary) $\omega$ is always {\it not} exact \cite{Fomenko}.)

Suppose that the symplectic form is exact:
$\omega = d\gamma,$ where $\gamma$ is some 1-form
(for classical systems $\gamma=p_{i}dq^{i}$). (Even
if the symplectic form is not globally exact, such 1-forms always exist locally.) Then the matrix
$\omega_{\mu\nu}$ can be
expressed in terms of the components of $\gamma=A_\mu
dx^{\,\mu}$:
$\omega_{\mu\nu} = \partial_\mu A_\nu -
\partial_\nu A_\mu.$
And the action can be written as follows \cite{alisa}:
\begin{equation}
\label{exact}
S=\int_{x_1}^{x_2}\gamma - \int_{t_1}^{t_2}Hdt= \int_{t_1}^{t_2} \Big( A_\mu \dot{x}^{\,\mu} - H
\Big) dt.
\end{equation}
Actually, the action (\ref{exact})  is just a special case of
Birkhoff theory (Pfaff-Birkhoff variational principle) \cite{McEwan,Birk,Sant}
$$ S= \int_{t_1}^{t_2} \Big( A_\mu (x,t) \dot{x}^{\,\mu} + B(x,t)
\Big) dt$$
with
$B$ staying for Hamiltonian. The equations of motion should be derived
from the variational principle with Dirichlet boundary conditions.
This formalism is valid
also for non-conservative systems with time-dependent
symplectic structure.

The
variation of the action
(\ref{exact}) yields
Hamiltonain equations of motion if one imposes Dirichlet
boundary conditions as it was done above in ${\mathbb R}^{2n}$ 
with the same problems persisting. In principle,
provided that we know an explicit form of $A_{\mu}$ we can introduce an
analogue of the functions $\phi$. But if we want to allow for different possible
choices of $\gamma$ then we are forced to admit the whole set of boundary conditions,
$\delta q(t_1)\!=\!\delta q(t_2)\!=\!\delta p(t_1)\!=\!\delta
p(t_2)\!=\!\!0$.

This mismatch between the number of boundary conditions for the variational
problem and the number of boundary conditions for the equations of motion often
makes physicists to worry about more that it really deserves. The habit to
identify one kind of boundary conditions with another is so strong that many distinguished
authors \cite{LL,Konop,HeTe} do not fix initial and final points for the action (\ref{action})
completely but impose only one half of the conditions ($\delta q(t_1)\!=\!\delta q(t_2)\!=0$),
even if they are going to discuss canonical transformations in the same text.

Actually, the story of this approach goes back to the early days of
Hamiltonian mechanics. But already in the book by Poincar{\' e} \cite{HP} a
discussion of canonical transformations appeared with the conclusion that
such principle has different forms in different canonical coordinates (in the sense
of different boundary conditions). For this reason it is commonly accepted in
mathematical literature to fix the boundary points completely by $4n$ conditions
\cite{AKN}, thus providing the variational principle with a kind of invariance.
And it is pleasant to note that the same is done in the classical textbook for physicists
\cite{Gold} (and in some other physical books \cite{Fliess}) where it is also stressed that only 
the whole set of boudary conditions
allows one to add a total time derivative term $\frac{d}{dt}f(p,q)$ to the
integrand in (\ref{action}).
\\

{\LARGE \section {The case of ``wrong'' boundary conditions}}
In this Section we would like to add several comments and to give a
variational characteristic of correctly chosen boundary points, which is 
in a sense obvious and presumably
not easy to use but, may be, worth mentioning still. Some other problems
would be discussed in a forthcoming article by L.V. Prokhorov and A.S. Ushakov.

In any case the correct statement is that the
physical trajectory brings the action to the stationary value in the class of
trajectories with fixed boundary points in the phase space.
But an important fact about the variational principles is that
one would probably wish to use the relevant principle in order to
get the equations of motion or to apply it instead of the
equations if it can make the mathematics easier. At this point we
encounter with a kind of problem: the number of boundary
conditions required is greater than the number of Hamiltonian
equations of motion. And for every initial point at $t_1$ there is
only one final point at $t_2$ for which the extremizing trajectory
exists. To get the physical trajectory one needs to guess the
unique final point from the continuum of all the phase space
points. But this problem can't lead to a false trajectory because
if the final point is chosen wrong the equations of motion can not
be satisfied and the action has no critical point in our class of
variations. One could even say that it is not a problem at all because one
can choose arbitrary initial point and fix the final point
formally without any idea of where it is fixed \cite{alisa}; it will allow him
to get the equations of motion and find the location of the final
point after that.

The well-known and very important property of the action (\ref{action}) is that it is bounded neither
from below nor from above. As a consequence, the physical trajectory is not a kind
of extremum but rather is a saddle point. In particular, it means that
the direct methods of calculus of variations can not be directly applied. And it
was not until about 30 years ago that certain progress has been made in this direction 
\cite{Rab1,Rab2,CoZeh,HoZeh} due to technique of approximation by some
finite-dimensional critical point problem \cite{Rab1}. It means also that if one shifts
a final point a little bit from its correct position then the range of the action does not change
radically, just its saddle point is gone.

We can regard the first variation of the action (\ref{action}) as a linear operator
acting from $L^2$-space of functions $\delta x(t):\ [t_1,t_2]\rightarrow{\mathbb R}^{2n}$
to ${\mathbb R}$, which depends on a chosen path from $x(t_1)$ to $x(t_2)$. Its operator norm
equals $||{\hat \delta}_S||=
\sqrt{\int\limits_{t_1}^{t_2}dt\sum_i\left(\left(\dot q_i-\frac{\partial H}{\partial p_i}\right)^2
+\left(\dot p_i+\frac{\partial H}{\partial q_i}\right)^2\right)}$. And only for the physical trajectory we
have $||{\hat \delta}_S||=0$, otherwise $||{\hat \delta}_S||>0$. If we have a ``wrong'' final point
then for all the possible paths $||{\hat \delta}_S||>0$. Moreover, the minimal value of
$||{\hat \delta}_S||$ normally exists, so that it is separated from $0$.

The minimal path for the functional $||{\hat \delta}_S||$ is given by
\begin{equation*}
\frac{d}{dt}\left(\dot q_i-\frac{\partial H}{\partial p_i}\right)=
-\frac{\partial^2 H}{\partial q_i\partial p_k}\left(\dot q_k-\frac{\partial H}{\partial p_k}\right)+
\frac{\partial^2 H}{\partial q_i\partial q_k}\left(\dot p_k+\frac{\partial H}{\partial q_k}\right),
\end{equation*}
\begin{equation*}
\frac{d}{dt}\left(\dot p_i+\frac{\partial H}{\partial q_i}\right)=
\frac{\partial^2 H}{\partial q_i\partial p_k}\left(\dot p_k+\frac{\partial H}{\partial q_k}\right)-
\frac{\partial^2 H}{\partial p_i\partial p_k}\left(\dot q_k-\frac{\partial H}{\partial p_k}\right)
\end{equation*}
which allows exactly $4n$ independent boundary conditions. If these conditions are
``right'' we have also $\dot q_i-\frac{\partial H}{\partial p_i}=0$,\ 
$\dot p_i+\frac{\partial H}{\partial q_i}=0$ and $||{\hat \delta}_S||=0$. In other cases
$||{\hat \delta}_S||>0$. Let us consider a very simple example with $n=1$ and $H=\frac{p^2}{2}$.
It implies $||{\hat \delta}_S||^2=\int\limits_{t_1}^{t_2}dt\left((\dot q-p)^2+{\dot p}^2\right)$. The
minimum is given by $q(t)=\frac{C_1}{6}t^3+\frac{C_2}{2}t^2+(C_3-C_1)t+C_4$ and 
$p(t)=\frac{C_1}{6}t^2+C_2 t+C_3$. For the initial point $(q_0,p_0\neq0)$ we have
to choose $(q_0+p_0(t_2-t_1),p_0)$ as the final point. But if we take
$(q_0+\alpha p_0(t_2-t_1),p_0+\beta)$, the constants $C_i$ would be $C_4=q_0$,
$C_3=p_0$, $C_1=\frac{6\beta+12(\alpha-1) p_0}{(t_2-t_1)^2+6(t_2-t_1)}$,
$C_2=\frac{\beta}{t_2-t_1}-\frac{C_1(t_2-t_1)}{2}$ and finally $||{\hat \delta}_S||^2=
C_1^2\left(\frac{7}{12}(t_2-t_1)^3+(t_2-t_1)\right)+\frac{\beta^2}{t_2-t_1}$. 
Only for $\alpha-1=\beta=0$ do we have
$||{\hat \delta}_S||=0$.

Of course, we used here even more complicated equations of motion than the original ones.
But the variational problem is quite different from that of (\ref{action}) because
$||{\hat \delta}_S||$ has a unique global minimum, and its value in principle can be
found by direct methods of variational calculus. After that, these minimal values can
be considered as a function of the final point (the initial point is fixed) which, in turn,
also has a unique global minimum in the ``right'' final point. The minimal value is zero.

Note that if the symplectic manifold is different from ${\mathbb R}^{2n}$ we may consider the
variations as elements of the tangent space. We don't want to go into any details here
but for any symplectic manifold there exist (not unique) an almost complex structure and
a Riemannian metric which are in a sense compatible with $\omega$ (see, for example, \cite{HoZeh},
p. 14), and it makes a room for the constructions explained above to be used in the
general setting of the action (\ref{exact}).
\\

\begin{center}
\begin{figure}
\includegraphics[width=7cm, keepaspectratio]{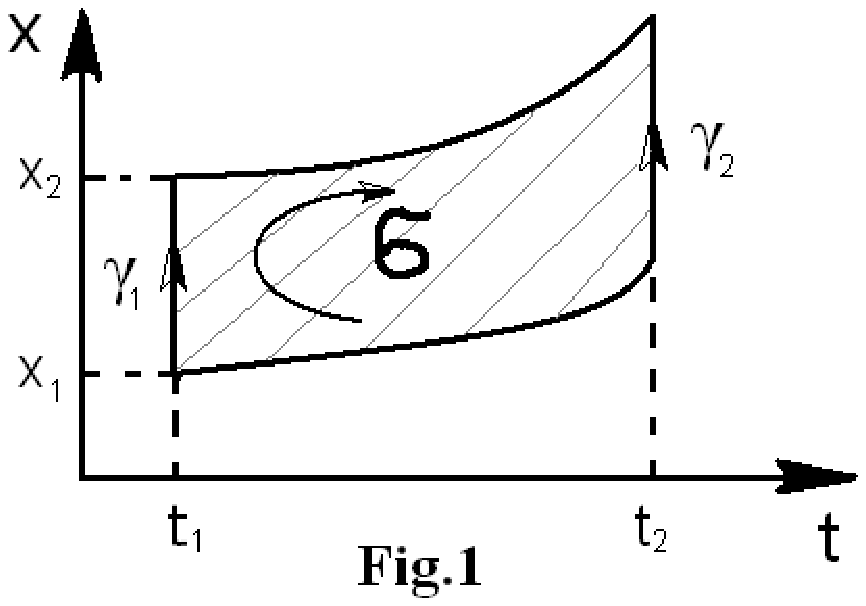}
\end{figure}
\end{center}

{\LARGE \section {Invariant variational principle}}

In the general case one should use the symplectic form
$\omega$ explicitly. And all the vectors we have
are in the tangent space. It means that the surface of integration
should be 2-dimensional. Looking at the formulae (\ref{actprime}) and (\ref{exact}) 
we expect that the action should have the following form:
\begin{equation}
\label{invariant}
S= \int_{\sigma}\left (
\omega - dH \land dt\right )=\int_{\sigma}\left (
\frac{1}{2}\omega_{\mu\nu}dx^{\mu}\land
dx^{\nu} - dH \land dt\right ).
\end{equation}

More precisely, we consider a trivial bundle over $M^{2n}: F^{2n +
1}=M^{2n}\times \mathbb{R}$ (extended phase space) and endow it
with a new coordinate $t$ (time) so that the basis of 1-forms
gains one more, (2n+1)-th, element, $dt$, with an axiom $dt\land
dx=-dx\land dt$. The differential forms $\omega$ and $dH$ are
defined in $F$ simply by their coordinate expressions in $M$ which
may be invariantly interpreted as a pullback of these forms
generated by the natural projection of $F$ to $M$. Any
one-parameter family of initial points $x(\varepsilon, t_1),$
$\varepsilon \in [0, E]$ defines a two-dimensional surface
$\sigma$ of trajectories in $F$ with the curve of the final points at $t=t_2$.
It is important that the initial and
final curves ($\gamma_1$ and $\gamma_2$) should be transversal to
the physical trajectories for the surface $\sigma$ to be
well-defined. One possible way to ensure it is to choose the
curves transversal to the hypersurfaces of constant Hamiltonian.
In this case the parameter $\epsilon$ gains also a possible
interpretation as a value of the Hamiltonian function, so that one
considers a family of trajectories with different amounts of
energy. In the rest of the Section we prove under these assumptions
the following {\bf Theorem}:

{\it 1. The surface $\sigma$ is a stationary one for the
action (\ref{invariant}) in the
class of smooth surfaces $x=x(\varepsilon, t),\quad \varepsilon
\in [0,E], \ t \in [t_1,t_2]$ with fixed ends $\delta
x(\varepsilon,t_1)=\delta x(\varepsilon,t_2)=0,\quad
\forall\epsilon$ (Fig. 1).}

{\it 2. An arbitrary surface $\tilde\sigma$ from this class is stationary
if and only if its boundary trajectories $x(0,t)$ and $x(E,t)$ are physical, i.e.
they satisfy the equations of motion.}

Let us start with a simple case when the surface $\sigma$ and small variations of it
belong altogether to one coordinate chart of the phase space manifold. Strictly speaking, 
it means that we use only a contractible domain in $M^{2n}$ and there does exist
some suitable 1-form $\gamma$. From its definition $\omega=d \gamma$ it follows that the action
(\ref{invariant}) equals to the action (\ref{exact}) on $\partial\sigma$ and the statement
is trivial. Nevertheless we want to proceed with explicit calculations in order to make an illustration
of how it works if one does not know an appropriate 1-form $\gamma$.
We perform the variations here in a somewhat formal way. We vary coordinates even
under the differential symbols in differential forms as if they were just ordinary functions. It can be justified if we take into account that for the small variations of the surface there is a natural one-to-one correspondence between points of the initial and the final surfaces. It allows us to vary the integrand instead of the domain of integration. It is also important
to mention that we consider only continuous variations of the surface ${\sigma}$ so that all the surfaces
are homotopic to each other and the Stokes` theorem is applicable.

The variation of the action (\ref{invariant}) is equal to
\begin{multline*}
\delta S= \int_{\sigma}
\Big( \frac{1}{2}\partial_\alpha\omega_{\mu\nu}\delta
x^\alpha dx^{\mu} \land dx^{\nu} +
\frac{1}{2}\omega_{\mu\nu}\partial_\alpha (\delta x^{\mu})
dx^{\alpha} \land dx^{\nu} + \\
 + \frac{1}{2}\omega_{\mu\nu}\partial_\alpha (\delta x^\nu)\,
dx^{\mu} \land dx^{\alpha} -
\partial_\beta (\partial_\alpha H \delta x^{\alpha})dx^{\beta} \land dt\Big).
\end{multline*}
We notice that the second and third terms under the integral are
equal, and integration by parts gives
\begin{multline*}
\delta S= \int_{\sigma}
\Big(\frac{1}{2}\partial_\alpha\omega_{\mu\nu}\delta
x^\alpha\, dx^{\mu} \land dx^{\nu} +\\ + \big(
\partial_\alpha (\omega_{\mu\nu}\delta x^{\mu}) -
\partial_\alpha\omega_{\mu\nu}\delta x^{\mu}\big)
dx^{\alpha} \land dx^{\nu} - \\
- \partial_\beta (\partial_\alpha H \delta x^{\alpha})dx^{\beta} \land dt\Big).
\end{multline*}
Using the Jacobi identify
\begin{equation*} \partial_\alpha \omega_{\mu\nu} = - \partial_\nu
\omega_{\alpha\mu} - \partial_\mu \omega_{\nu\alpha} = -
\partial_\nu \omega_{\alpha\mu} + \partial_\mu
\omega_{\alpha\nu}
\end{equation*}
we get
\begin{multline*}
\delta S= \int_{\sigma}
\Big(\frac{1}{2}\partial_\mu\omega_{\alpha\nu}\delta
x^\alpha dx^{\mu} \land dx^{\nu} -
 \frac{1}{2}\partial_\nu\omega_{\alpha\mu}\delta x^\alpha
dx^{\mu} \land dx^{\nu} +\\
+\big(
\partial_\alpha (\omega_{\mu\nu}\delta x^{\mu}) -
\partial_\alpha\omega_{\mu\nu}\delta x^{\mu}\big)
dx^{\alpha} \land dx^{\nu} -\\- \partial_\beta
(\partial_\alpha H \delta x^{\alpha})dx^{\beta} \land
dt\Big) = \\
= \int_{\sigma}
\Big( - \frac{1}{2}\partial_\mu\omega_{\alpha\nu}\delta
x^\alpha dx^{\mu} \land dx^{\nu}  +
\frac{1}{2}\partial_\nu\omega_{\alpha\mu}\delta x^\alpha
dx^{\nu} \land dx^{\mu} +\\
+ \partial_\alpha (\omega_{\mu\nu}\delta x^{\mu})
dx^{\alpha} \land dx^{\nu} - \partial_\beta
(\partial_\alpha H \delta x^{\alpha})dx^{\beta} \land
dt\Big) =  \\
= \int_{\sigma}
 \Big(\frac{1}{2}\partial_\alpha (\omega_{\mu\nu}\delta x^{\mu})
dx^{\alpha} \land dx^{\nu} -  \partial_\alpha
(\partial_\beta H \delta x^{\beta})dx^{\alpha} \land
dt\Big) =
\end{multline*}
\begin{multline*}
= \int_{\partial\sigma}
 \Big(\omega_{\mu\nu}\delta x^{\mu}dx^{\nu} -  \partial_\beta H \delta
x^{\beta}dt\Big) = \\
= \int_{\partial([0,E]\times[t_1,t_2])}
 \Big(\omega_{\mu\nu}\delta x^{\mu}dx^{\nu} -  \partial_\mu H \delta
x^{\mu}dt\Big) = \\
= \int_{t_1}^{t_2}
 \Big(\omega_{\mu\nu}\dot{x}^{\nu}(E,t) -  \partial_\mu H(x(E,t))\Big) \delta
x^{\mu}dt-\\
-\int_{t_1}^{t_2}
 \Big(\omega_{\mu\nu}\dot{x}^{\nu}(0,t) -  \partial_\mu H(x(0,t))\Big) \delta
x^{\mu}dt.
\end{multline*}
The variations $\delta x^{\mu}$ are arbitrary and can be performed independently
at $\epsilon=0$ and at $\epsilon=E$. Hence, $\delta
S=0$ is equivalent to Hamiltonian equations of motion
$\omega_{\mu\nu}\dot{x}^{\nu}=\partial_\mu H$ for $x(0,t)$ and $x(E,t)$.
It means that in the space of all possible one-parameter families of curves 
we have an ``infinite-dimensional
stationary manifold'' for the action (\ref{invariant}). The intermediate
trajectories do not influence the stationarity condition because 
the differential form under the integral in (\ref{invariant}) is closed and
any variation within one homotopy class with the whole
boundary $\partial\sigma$ fixed is subject to the Stokes' theorem.

It's worth to mention that before the last step of the calculation
all $x$ and $\delta x$ could be regarded as functions of point in
$F^{2n+1}$. Only at the last step we restrict ourselves to the
boundary lines and consider these functions as well-defined
functions of time. It means that we could vary the time in (\ref{invariant})
independently. It is easy to see that the consequence of such
variation is that the Hamiltonian does not change with time
(along the physical trajectory). It follows also from the
equations of motion. It's not surprising that we have got nothing
new because the variation of time is equivalent to some variation
of dynamical functions $x(t)$. 

Suppose now that the surface $\sigma$ in our calculation is divided into several parts by a number of 
internal lines which are transversal to each other and to $\partial\sigma$. 
For each part the variation of the action yields the equations of motion on its boundary.
Now we want to glue these parts together. How can the result for initial $\sigma$
be restored after that? Clearly we have to add up the integrals over all the parts and
demand that the variations of different parts should be equal each other on those boundaries 
which are going to be identified. Then integrals over all internal lines in $\delta S$ come up
twice with opposite signs and cancel each other. (We know that it really {\it had} to be so due 
to the Stokes' theorem.) Again we get the Hamiltonian equations of motion for the boundary
trajectories of $\sigma$. 
It allows us to prove the Theorem in its full generality.
Indeed, even if we can't embed a surface $\tilde\sigma$ into a contractible domain in $M^{2n}$, we still can
divide it into small parts such that every part together with its nearest neighbours belongs to some 
contractible domain. The action functional is invariant under coordinate transformations. 
Each part of the surface can be varied in 
any coordinate system and contributions of division lines cancel each other. The final result
contains only equations of motion on the boundary trajectories of $\tilde\sigma$.
The theorem is proven.

After this analytic derivation we would like to add a nice geometric picture of the theorem. As it was already mentioned, those variations which do not change boundary trajectories are irrelevant due to the Stokes' theorem. Let us consider a variation of the surface $\sigma$ which is not trivial only in a small vicinity of a part of one boundary trajectory. Then the variation is just the difference between two integrals over small pieces of $\sigma$ and $\tilde\sigma=\sigma+\delta\sigma$. And due to the Stokes' theorem again this difference is equal to the integral over some surface which connects the initial trajectory and its image after the variation. So, it means that the integral of $\omega-dH\land dt$ over arbitrary thin stripe along the physical boundary trajectory should be of order ${\cal O}((\delta x)^2)$ with no regard to the orientation of the stripe. (Note also that we can take a not too long part of the trajectory and use the Darboux coordinates on the stripe if we like.) It means that  $(\omega-dH\land dt)(l,a)=0$ for any vector $a$ and a vector $l$ tangent to the trajectory. It is precisely the equations of motion.

Let us also briefly mention that we could define a ``superextended`` phase space with {\it two} new coordinates,
$H$ and $t$. In this case we should perform variations only on a hypersurface $H=H(x)$ and may consider the parameter $\epsilon$ as a coordinate along the H-axis (Fig. 2).
\\

\begin{center}
\begin{figure}
\includegraphics[width=7cm, keepaspectratio]{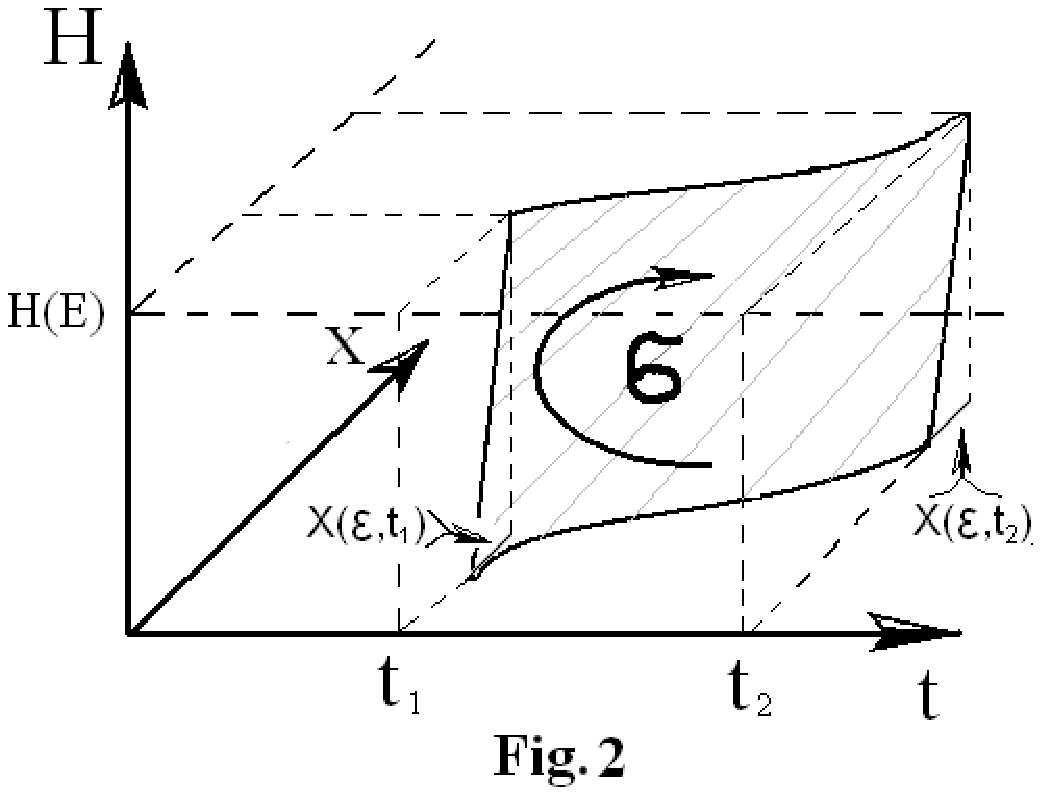}
\end{figure}
\end{center}

{\LARGE \section {Some additional remarks}}

Curiously enough, the idea of some 2-form integration over 2-dimensional surface between two trajectories appeared recently in \cite{kochan} which goes in a direction somewhat opposite to the lines of our work. Systems with velocity dependent forces, generally admiting neither Hamiltonian nor Lagrangian formulation (see, however, \cite{Git1} and
\cite{Git2}), 
are considered in this reference in the language of positions and velocities (we would like to remind here that, on the contrary, our main goal is the treatment of Hamiltonian systems for which it is generally impossible to separate coordinates and momenta). For these (generally dissipative) systems a variational principle is obtained which yields the equations of motion and some more equation on the bulk of the integration surface with unclear dynamical meaning. So that a difficult problem of joint solvability arises, see \cite{kochan} for details. In spite of certain similarity, the intersection of this principle with ours is rather trivial. In the case of conservative systems it reduces to the Lagrangian version of the action (\ref{invariant}), but the relevant systems are those for which the coordinates and momenta are clearly separated, and much simpler action principle of the form (\ref{action}) or (\ref{exact}) exists.

Note also that we could formulate the variational principle invariantly but without 
dealing too much with the language of exterior calculus.
The obvious relation $dx^{\mu}=\dot{x}^{\mu}dt + x^{\prime\mu}d\varepsilon$
leads us to the action integral:
$$S=\int_{t_1}^{t_2}\left(\int_{0}^{E}\omega_{\mu\nu}\dot{x^{\nu}}x^{\prime\mu}d\epsilon-
\left(H(x(E,t))-(H(x(0,t))\right)\right)dt.$$
And varying it with respect to $x^{\mu}(\epsilon,t)$ in the same way as in the
Section 4, we get the same result
$$\delta S= \int_{t_1}^{t_2}\Big(
  \omega_{\mu\nu}\dot{x}^{\nu} - \partial_\mu H \Big) \delta
  x^{\mu} \bigg|_{\varepsilon =E} dt-\int_{t_1}^{t_2}\Big(
  \omega_{\mu\nu}\dot{x}^{\nu} - \partial_\mu H \Big) \delta
  x^{\mu} \bigg|_{\varepsilon =0} dt.$$

Then we have to
mention the problem of boundary conditions again. It may seem to
be even more intricate in the case of non-exact forms because if
the initial points in the phase space are given one has to guess
properly the final points at least for {\it two} trajectories (at
$\epsilon = 0$ and $\epsilon=E$) for the stationary surfaces to exist. But
we can easily reformulate our principle: choose in $F^{2n+1}$ only
one physical trajectory and one auxiliary line with the same
initial and final points such that a nonsingular surface in the
phase space exists with the boundary equal to these two curves.
Then we can take our action integral (\ref{invariant}) along the surfaces of
that kind with the auxiliary line fixed and the physical
trajectory free to change (except the boundary points, of course).
By literally the same calculations as in the Section 4 it can be
easily verified that all surfaces for which the equations of
motion are valid on the trajectory would compose the stationary
manifold for the action considered.

And a final remark deals with Maupertius principle. Suppose we
consider only the first term in the action (\ref{invariant}) but vary this
action only in the class of surfaces with boundary trajectories laying on 
two different hypersurfaces
of constant Hamiltonian. The result is
that for every vector $l^{\nu}$ tangent to the $H=const$
hypersurface in $M^{2n}$ the equation $\omega_{\mu,\nu}\dot x^{\mu}l^{\nu}=0$
should be held true on the boundary trajectories. In
the phase space $M^{2n}$ this equation uniquely defines a curve
which is compatible with the equations of motion
$\omega(\dot x, .)=-dH(.)$
because d$H(l)=0$
for the vectors considered. But the time coordinate may be chosen
arbitrarily because the equation is reparametrization invariant.
So we get the invariant form of the Maupertius principle.\\

{\LARGE \section {Examples and discussion}}
The usual action principle for Hamiltonian systems is given by (\ref{exact})
(or by (\ref{action}) for classical systems). On exact manifolds the 1-form $\gamma$
is globally well-defined and the action (\ref{exact}) can be perfectly used, but
one should remember that although the choice of $\gamma$ is a coordinate-free procedure,
it is not unique and effectively means a kind of distinction between what we would call momenta and what
we would call coordinates. For non-exact manifolds $\gamma$  exists only locally, and
even if with a particular choice of $\gamma$ we were able to pursue a variational approach for
one trajectory it doesn't mean yet that we would be able to repeat it for some another
path without changing the set-up. 

Our variational principle is completely invariant and can be used for any Hamiltonian system.
Let us consider a few very simple examples. First of all, we take a sphere with
$\omega=\cos\theta\ d\theta\wedge d\varphi$ in spherical coordinates, $\theta\in[-\frac{\pi}{2},\frac{\pi}{2}]$,
$\varphi\in [0,2\pi)$ and Hamiltonian $H=\sin\theta$. (One can check that $\omega$ is non-degenerate
and $H$ is smooth near the poles $\theta=\pm\frac{\pi}{2}$, for example, by going to new
coordinates $x=\cos\theta\cdot\cos\varphi$, $y=\cos\theta\cdot\sin\varphi$ in which
$\omega=\frac{dy\wedge dx}{\sqrt{1-x^2-y^2}}$ and $H=\sqrt{1-x^2-y^2}$ are obviously well-defined
near $x=y=0$.) The equations of motion are $\dot\theta=0$, $\dot\varphi=1$. Let $\varphi(\epsilon,t_1)=
0$, $\theta(\epsilon,t_1)=\epsilon$, $0\leq\epsilon\leq E<\frac{\pi}{2}$ be a family of initial data and
$\varphi(\epsilon,t_2)=\alpha (t_2-t_1)$, $\theta(\epsilon,t_2)=\epsilon$ -- a family of final points.
(We use $\alpha$ to show what happens if final points are ''wrong''.) For this case (one degree of freedom) irrelevance of intermediate trajectories is obvious because the surface of integration $x(\epsilon ,t)$ is completely defined by boundary paths $\theta (\phi (t))$ at $\epsilon =0$ and $\epsilon =E$:\ $\varphi(E,t)$, $\varphi(0,t)$, $\theta(E,t)=
f_2(\varphi(E,t))$, $\theta(0,t)=
f_1(\varphi(0,t))$. The action (\ref{invariant}), after one half of the integrations have been performed, 
turns into the following form:
\begin{multline*}
S=\int\limits_{0}^{\alpha(t_2-t_1)}d\varphi \left(\sin (f_2(\varphi))-\sin (f_1(\varphi))\right)-\\
-\int\limits_{t_1}^{t_2}dt \left(\sin (f_2(\varphi(E,t)))-\sin (f_1(\varphi(0,t)))\right).
\end{multline*}
The variations of $f_2(\varphi), \varphi(E,t)$ and $f_1(\varphi), \varphi(0,t)$ should be
performed independently, and actually we can even fix one of the boundary trajectories and obtain the
equations of motion only for the second path. If we vary $\varphi(E,t)$ as a function of time
we have $\delta_{\varphi}S=-\int dt\cos(f_2(\varphi))f_{2}^{\prime}\ \delta\varphi$. It implies
$f_2=const$, i.e. $\dot\theta=0$. Now we vary $f_2(\varphi)$ independently:
$$\delta_{f_2}S=\int\limits_{0}^{\alpha(t_2-t_1)}d\varphi\ \cos (f_2(\varphi))\delta f_2(\varphi)
-\int\limits_{t_1}^{t_2}dt\ \cos (f_2(\varphi))\delta f_2(\varphi)$$
and after changing the variable in the first integral according to $d\varphi=\dot\varphi dt$
we get $\dot\varphi=1$. The consistency condition ($\Delta\varphi=\int\dot\varphi dt$) leads
to $\alpha=1$. There is no stationary point for the action $S$ otherwise.

Actually, ${\mathcal S}^2$ is a very simple manifold and one could use 1-form
$\gamma=\sin\theta\ d\varphi$ on it. But this $\gamma$ is singular in the poles, and if
we were not so clever we could take $\tilde\gamma$ with poles on the physical trajectory.
One more remark is that we could make $E\to\frac{\pi}{2}$ and convert the action (\ref{exact})
to a rather nice form
\begin{equation}
\label{HZ}
S=\int\limits_{\mathcal D}\omega-\int\limits_{t_1}^{t_2}Hdt
\end{equation}
where $t_2-t_1=2\pi$ and $\mathcal D$ is the upper part of the sphere with a periodic
trajectory as a boundary. This expression is valid only for contractible closed trajectories
and is being successfully used for the purposes of symplectic topology \cite{HoZeh,Floer}, but 
it is not well known among physicists. (The Authors were unaware of it before searching
the literature for historical references for this article.) We have to note that the actions
(\ref{invariant}) and (\ref{HZ}) present even more intricate problems for the direct variational methods
because if one changes the homotopy class of $\sigma$ or $\mathcal D$ he will not affect 
the stationarity of the action with respect to small continuous variations of the surface
but will shift the value of the action by some element of the so-called period group which
consists of all the real numbers obtained by integrating $\omega$ over submanifolds
homeomorphic to ${\mathcal S}^2$. Sometimes this ambiguity is only of academic interest because, for example,
for systems on ${\mathcal S}^2$ it means only that we could force the disk $\mathcal D$
to wrap several times around the whole phase space. 
(Note that in this case the values of action (\ref{HZ}) would aquire an additional constant,
a multiple of the total ${\mathcal S}^2$ area. It will change nothing for the variations of
(\ref{HZ}) but no 1-form $\gamma$ would be correctly defined.)
But unfortunately it is not always so
good, and sometimes the period group may even be everywhere dense in $\mathbb R$, see
\cite{HoZeh}, p. 228.

Needless to say, our principle works equally well for both contractible
and non-contractible loops. Consider now a torus ${\mathcal T}^2$ with $\omega=d\theta\wedge d\varphi$,
$\theta\in [0,2\pi)$, $\varphi\in [0,2\pi)$, $H=\sin\theta$ and equations of motion
$\dot\theta=0$, $\dot\varphi=\cos\theta$. (This example is oversimplified not only
due to the local existence of $\gamma=\theta\ d\varphi$ but also because for the torus
one can use a trick \cite{CoZeh} of considering the action (\ref{actprime}) for the periodic
functions in ${\mathbb R}^2$. But still we do not want to complicate it here.) Our periodic boundary conditions
$\theta_i=\theta_f=\epsilon\in[0,E]$, $\varphi_i=0$, $\varphi_f=2\pi n$, $n\in{\mathbb Z}$ imply
$$S=\int\limits_{0}^{2\pi n}d\varphi \left(f_2(\varphi)-f_1(\varphi)\right)
-\int\limits_{0}^{T}dt \left(\sin (f_2(\varphi(E,t)))-\sin (f_1(\varphi(0,t)))\right)$$
Suppose that we fix the $\epsilon=0$ path and vary only the $\epsilon=E$ trajectory. Variation
of $f_2(\varphi)$ yields $f_2=const$ and after that varying $\varphi(t)$ we get $\dot\varphi=\cos\theta$
with the consistency condition again $\Delta\varphi=\int\dot\varphi dt$. The last equation means that
trajectories satisfying the periodicity relation $\varphi(t+T)=\varphi(t)+2\pi n$ exist
at $\cos\theta=\frac{2\pi n}{T}$. There are several different solutions to the
equation $\delta S=0$ for large enough $T$. (Different solutions have different boundary 
trajectories, and each one of these solutions
also has an infinite degeneracy due to possibility of wrapping $\sigma$ any number of times
around the torus without changing the boundaries.) In the invariant setting we do not have to worry about
the location of these stripes $\sigma$, but if we decided to use $\gamma=\theta\ d\varphi$ it would be necessary to
cut the torus along the line of $\theta =0$.

These $n=1$ examples do not allow us to illustrate the Maupertius principle. For this purpose let's consider
the simplest example of $n=2$ system: $M^{2n}={\mathcal S}^2\times{\mathbb R}^2$,
$\omega=\cos\theta\ d\theta\wedge d\varphi+dp\wedge dq$, $H=\sin\theta+\frac{p^2}{2}$. The
equations of motion are $\dot\theta=0$, $\dot p=0$, $\dot\varphi=1$, $\dot q=p$. To apply the
Maupertius principle we first find the surface of constant Hamiltonian: $\sin\theta+\frac{p^2}{2}=const$.
After that we take the boundary conditions: $\varphi_i=0$, $\varphi_f=t_2-t_1$, $\theta_i=\theta_f=\alpha\epsilon$,
$q_i=0$, $q_f=\beta\epsilon(t_2-t_1)$, $p_i=p_f=\beta\epsilon$ and the abbreviated action
$$S_{M}=\int\omega=\int\limits_{0}^{t_2-t_1}d\varphi \left(\sin (f_2(\varphi))-\sin (f_1(\varphi))\right)+
\int\limits_{0}^{\beta\epsilon(t_2-t_1)}dq\left(g_2(q)-g_1(q)\right)$$
where $\theta(E,t)=
f_2(\varphi(E,t))$, $p(E,t)=
g_2(q(E,t))$ and similarly for $f_1, g_1$. Varying the second trajectory we have
$\delta S_{M}=\int d\varphi\ \cos(f_2)\ \delta f_2+\int dq\ \delta g_2$. We consider $q=q(\varphi)$ and
use the relation $\cos(f_2)\ \delta f_2+g_2\delta g_2=0$ on the $H=const$ surface. The result
is $\delta S_{M}=\int d\varphi\ (-g_2+q^{\prime}(\varphi))\ \delta g_2$ and $q^{\prime}(\varphi)=p$.
Independent variation of $q(\varphi)$ yields $\int d\varphi\ g_2\delta q^{\prime}$ and $g_2=const$.
Due to $H=const$ it implies $f_2=const$ and finally we have $\theta=\alpha E=const$, 
$p=\beta E=const$, $\frac{q}{\varphi}=\beta E=const$.

We have presented a new completely invariant approach to the variational formulation of
Hamiltonian mechanics. Our principle can be applied to any Hamiltonian system,
but for the direct methods of variational calculus it inherits all the usual mathematical difficulties
related to action principles in Hamiltonian form. Therefore the question of how far one could
go further with it remains an open problem of mathematical nature.

The Authors are grateful to anonymous referees for useful comments and suggestions.
\\

\end{document}